\documentclass[twocolumn,showkeywords,preprintnumbers,amsmath,amssymb]{revtex4}

\usepackage{graphicx}
\usepackage{dcolumn}
\usepackage{bm}

\begin{document}

\title{Quantum Measurement and Observable Universe}

\author{Daegene Song}

\affiliation{%
Department of Management Information Systems, Chungbuk National University, Cheongju, Chungbuk 362-763, Korea
}%

\date{\today}

\begin{abstract}
In this paper, we discuss that an observable-based single-system 
Copenhagen and entanglement-based two-system von Neumann measurement 
protocols in quantum theory can be made equivalent by considering 
the second part of the two-system scheme to be a Dirac-type 
negative sea filling up the first system. Based on this equivalence, 
and by considering the universe as a computational process, 
the choice of the apparatus state in the two-system protocol 
can be identified with the choice of the observable in the 
single-system scheme as negative sea filling up the observable universe. 
In particular, the measuring party's state is considered to be 
evolving backwards in time to the big bang as a nondeterministic 
computational process, which chooses the acceptable path as a 
time-reversal process of irreversible computation. The suggested 
model proposes that the prepared microstate of the universe, or reality, 
corresponds to the observer's choice, therefore, subjective reality. 
Thus, this effectively provides a specific description of the subjective 
universe model previously proposed, which is based on the 
symmetry breakdown between the Schr\"odinger and the Heisenberg pictures of quantum theory.

\end{abstract}

\maketitle
\section{Objectivity vs. Subjectivity}
\begin{figure}
\begin{center}
{\includegraphics[scale=.65]{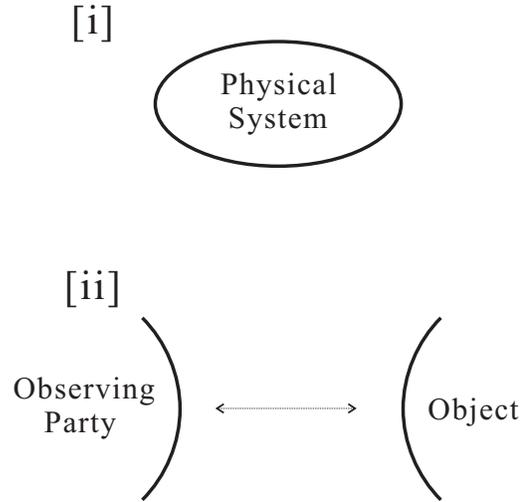}}

\end{center}
\caption{ [i] Objectivity: Previously, it was assumed that physics attempted to reveal objective reality. 
[ii] Subjectivity: With development of quantum theory, quantum information science in particular, 
the relationship between the observing party and the object was pursued.  Moreover, it has been claimed that 
subjectivity is quite natural considering that science is based on observation and experience.  
  }
\label{Subj}\end{figure}

Since John Bell's intuitive view of the verification of nonlocality in 
quantum theory \cite{bell} and other subsequent experiments \cite{aspect,tittel,experiment}, there has been 
resurgence in interest in the field of quantum foundations. 
Moreover, with various practical applications of quantum information, 
such as quantum computation \cite{deutsch}, quantum key distribution \cite{BB84}, and quantum 
communication \cite{cleve} (see \cite{nielsen} for a review), there has been an active 
discussion in regard to the very nature of counterintuitive 
quantum theory. One of the central issues in quantum foundations 
is the unusual characteristic of the measurement process, i.e., 
in the standard interpretation of quantum theory, it assumes 
a special status for the observing party. This subjective nature 
in quantum theory is problematic, since it has often been assumed that 
physics is a scientific study designed to come up with an objective 
pattern of physical systems, i.e., 
something that holds whether there is an observer or not and 
irrespective of who is observing it (Fig. \ref{Subj} [i]).

Although standard quantum theory deals with both unitary transformation and measurement, 
the latter has often been minimized. In particular, the measurement 
sector of quantum theory is often considered a more philosophical issue. 
However, the development of quantum information science began to 
incorporate the observer's role, i.e., the measurement part, into 
the core part of the theory, and it began to shift toward the relationship 
between the observing party and the object (Fig. \ref{Subj} [ii]). In fact, 
there has been discussion that the subjectivity of quantum mechanics 
may not be so strange after all. For example, it has been argued \cite{song2} that, 
unlike the premise that physics should pursue objective reality \cite{EPR}, 
subjectivity ought to be related to the limit of scientific knowledge. 
That is, the subjective description of standard quantum theory, with its 
state vector and observables, is a natural description of nature when we 
consider that physics could provide only a subjective observation of nature 
(Fig. \ref{Subj} [ii]). The subjectivity in quantum theory 
was in some sense inevitable as science progressed to reach this limit \cite{song2}.

\section{Symmetry Breakdown in Two-Picture Quantum Theory}
Motivated by the assumption that physics 
provides a description between the observer and the object, it was argued \cite{song5}
that the observable should be identified as a reference frame when observing the state vector. 
In particular, it was noted that just as the state 
vector was an exact and ultimate mathematical description of the 
object, i.e., when we accept the noncausal probabilistic 
nature at the fundamental level, the observable should 
also be an exact and ultimate mathematical description of the observer. 
Note that taking the observable as the exact  
mathematical description of the observer's state 
can be useful because there has been very 
active research in brain science and neural network 
approaches to identify the status of one's mental state, 
but it came up with only an approximation at best. 
However, the proposed approach claims to provide an exact 
and full mathematical description of the observer's state, 
at least when the observer is observing the given state vector.

Based on this assumption, it was shown that the two-picture formulation 
of quantum theory, namely, the Schr\"odinger and the Heisenberg pictures, 
runs into a contradiction when we consider the case of the reference 
frame also being the object to be observed, i.e., consciousness. 
This result is particularly interesting because recent rapid developments  
in the field of artificial intelligence (AI) have yielded 
 divided opinions about its limit, i.e., whether a 
computing device can eventually become self-aware, a term also known as 
strong AI. 
The result in \cite{song3} has shown that self-referencing consciousness is 
different from the rest at the fundamental level, i.e., it could not be computed by 
either classical or quantum computing systems.

On the other hand, as a resolution to the seemingly contradictory nature of consciousness causing a symmetry breakdown 
in the usual two-picture formulation,  
it was argued \cite{song2} that it is not okay to consider the observer separate from the object being observed.  
That is, not only does physics provide a subjective 
description between the observer and the object (see Fig. \ref{Subj} [ii]), but
 existence should also be subjective, i.e., the whole universe is not separable from the observer.  
It is a goal of this paper to address a specific description of 
what is meant by \lq\lq the whole universe\rq\rq \, being inseparable from the observer.

\section{Equivalence of Two Measurement Protocols}
\begin{figure}
\begin{center}
{\includegraphics[scale=.55]{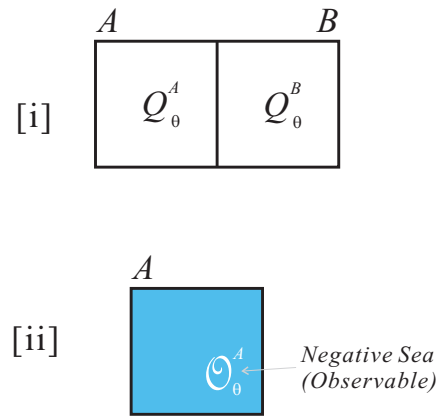}}

\end{center}
\caption{ [i] Entanglement-based two-system measurement protocol where $A$ and $B$ 
correspond to a system and an apparatus, respectively. 
[ii] Single-system Copenhagen measurement protocol in quantum theory with state vector and the observable. 
Equating the two measurement protocols could be done 
by considering the second system $B$ as a negative sea filling up system $A$ \cite{song1}.   }
\label{Dirac}\end{figure}

In quantum theory, there are two quite different measurement protocols. 
The first is a single-system protocol (Fig. \ref{Dirac} [ii])  
with a state vector and observable found in standard quantum theory. 
On the other hand, there is another protocol (Fig. \ref{Dirac} [i]) 
that introduces a second system of apparatus state and entanglement, which is used between the two systems.  
These two measurement 
protocols in quantum theory are troublesome when we consider the system of the whole universe. 
The second protocol seems to treat the observing party as a physical system and employ   
an objective view of the whole universe (Fig. \ref{Subj} [i]).  
However, the first single-system protocol 
gives the observing party a special status and tries to provide a description 
between the observing party and the object (Fig. \ref{Subj} [ii]).  
While the two-system scheme is in line with the objective law, the single-system protocol is in line 
with the subjective nature of science discussed previously.

In \cite{song1}, the equivalence of the two different measurement protocols is discussed.  
By considering the apparatus state of the entanglement-based two-system 
protocol (Fig. \ref{Dirac} [i]) as a Dirac-type negative sea filling 
up the first system, it should correspond to the observable 
in the single-system Copenhagen measurement protocol (Fig. \ref{Dirac} [ii]).  
This equivalence was applied to black hole radiation where there are
 two different kinds of protocols: entanglement-based two-system \cite{unruh} 
and observable-based single-system \cite{hawking1}. 
It was argued that the second system or inside of the black hole, i.e., $B$,  
should be considered a negative sea filling 
up the outside of the black hole horizon, i.e., $A$, and 
should be identified with the choice of observable, which is 
an ingredient of the first type of quantum measurement protocol, i.e., 
\begin{equation}
\mathcal{O}_{\theta}^{A} = Q_{\theta}^{B}
\label{TheEquation}\end{equation}
where $\theta$ is a number between $1$ and the number of equally accessible microstates. 
In particular, the relation in (\ref{TheEquation}) indicates that the choice of 
apparatus state, $\theta$, inside the black hole is in fact the 
choice of the observable outside the horizon.  On the 
one hand, this provides a resolution to the information loss problem 
in black hole radiation by considering it to be a quantum measurement process.  
On the other hand, 
it shows how an observer represented by a physical system of 
apparatus, i.e., in line with objectivity (Fig. \ref{Subj} [i]),
 gives rise to a special status in the standard single-system 
view of the measuring party, i.e., subjective description of nature (Fig. \ref{Subj} [ii] and 
Fig. \ref{Dirac} [ii]).

\section{Nondeterministic Computation and Negative Sea}
In this section, we apply the identification of the  
two quantum measurement protocols in (\ref{TheEquation}) to 
the cosmological model. 
In order to do so, we consider an observable-based single-system case, 
similar to the Copenhagen measurement protocol (Fig. \ref{Dirac} [ii]). 
Let $\Omega$ be the number of equally accessible microstates corresponding to the observable 
universe, i.e., outside the cosmic event horizon, for a given observer. 
Thus, the entropy of the observable universe corresponds to the usual Boltzmann's formula, i.e., 
$S=k \ln \Omega$ where $k$ is the Boltzmann's constant. 
It is noted that the entropy of the observable universe has been estimated in \cite{egan} using the Bekenstein-Hawking 
formula \cite{bekenstein,hawking,gibbons} (see \cite{bradford} for a review). 
Let us continue with a two-system case similar to the von Neumann protocol (Fig. \ref{Dirac} [i]) and 
consider a microcanonical state of the cosmological model 
as follows, 
\begin{equation}
|\Psi\rangle_{AB} = \frac{1}{\sqrt{\Omega}} \sum_{\theta =0}^{\Omega-1} |\theta\rangle_A |\theta\rangle_B
\label{Psi}\end{equation}
where system $A$ corresponds to the observable universe. 
Note that since $|\theta\rangle_A$ corresponds to degenerate state, there exists an observable, which we will 
call $\mathcal{O}_0^A$, such that $_{A}\langle 0|\mathcal{O}_0^A |0\rangle_A =$  
$_{A}\langle 1 |\mathcal{O}_0^A |1\rangle_A = \cdots  = $    $_{A}\langle \Omega-1|\mathcal{O}_0^A |\Omega-1\rangle_A $.   
Due to symmetry, there is also an equal number of observable choices, 
which we denote as $\mathcal{O}_{\theta}^{A}$ 
where $0\leq \theta \leq \Omega-1$, i.e., there are $\Omega$ degenerate observables, such that 
$_{A}\langle 0|\mathcal{O}_0^A |0\rangle_A =$  
$_{A}\langle 0 |\mathcal{O}_1^A |0\rangle_A =\cdots  =$  $ _{A}\langle 0|\mathcal{O}_{\Omega-1}^A |0\rangle_A $.

\begin{figure}
\begin{center}
{\includegraphics[scale=0.75]{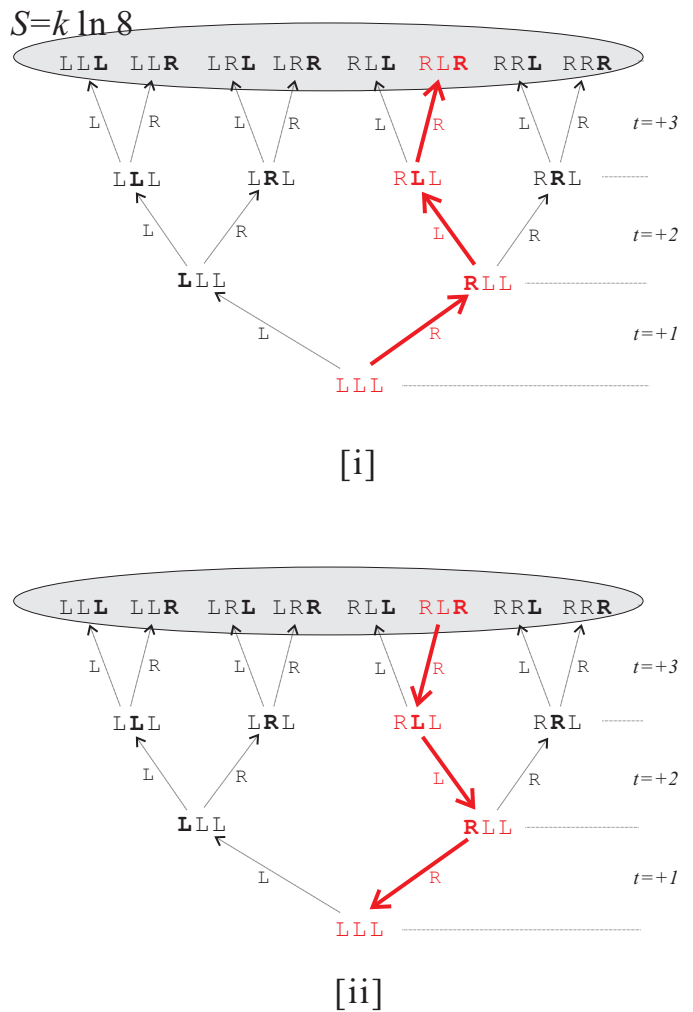}}

\end{center}
\caption{ Paths of [i] irreversible and [ii] nondeterministic computation.   
In [i], the paths of irreversible gates acting 
on three bits are shown.  After three logic gates are applied, 
the final state after $t=+3$ has the entropy 
$S=k\ln 8$.  The time-reversal process of irreversible gates, i.e., nondeterministic computation,  
is indicated in [ii].  }
\label{Irreversible}\end{figure}

We now wish to discuss how the apparatus system $B$ in (\ref{Psi}) can be related to 
the observable $\mathcal{O}_{\theta}^{A}$ in the 
single system case by using nondeterministic computation. 
In order to do so, let us first review the 
irreversible process of computation. Irreversible 
computation refers to a process where, given the output, 
it is not possible to indicate a single input, i.e., 
either the initial state of $a_1$ or $a_2$ may end up with $b$. 
For example, {\tt{restore-to-zero}} is a gate that sets the input, 
whether {\tt{0}} or {\tt{1}}, to {\tt{0}}.
When this gate is performed, it would not be possible 
to determine whether the previous state was ${\tt{0}}$ or ${\tt{1}}$ 
given the state ${\tt{0}}$.  It was shown by 
Landauer \cite{landauer} that in the case of a bit with equal probability 
the irreversible gate necessarily generates the increase of entropy by $k\ln 2$. 
On the other hand, nondeterministic computation corresponds to the case where,
 with the initial state $a$, the transition may be more than one, i.e., 
 \begin{equation}
 a \rightarrow \{ b_1, b_2\}
\label{nondeterministic}\end{equation}
 Indeed, nondeterministic computation may be considered to be 
a time-reversal of the irreversible computation.

It should be noted that nondeterministic computation 
is different from probabilistic computation, which 
has more than one path of computation with an  
assigned probability for each.  In the case of nondeterministic 
computation, it chooses the path from among more than one possibility  
that is acceptable.  
For example, let us take an example with the irreversible gate {\tt{restore-to-zero}} 
acting on three bits that are initially randomly prepared. 
As discussed earlier, this particular logic gate acts whether the input is 
{\tt{0}} or {\tt{1}} and sends it to {\tt{0}}. 
At $t=+1$, the first irreversible gate is applied to the first bit, 
which we will assume to be at {\tt{1}} rather than {\tt{0}}, and sends it to {\tt{0}}. 
The second bit is at {\tt{0}} and the {\tt{restore-to-zero}}
 gate acts on it to keep it at {\tt{0}} at $t=+2$.
Finally, at $t=+3$, the irreversible gate is applied 
to the third bit, which we will assume to be {\tt{1}}, 
and restores it to {\tt{0}}.

If we make a diagram 
indicating the path of computation so that the irreversible gate takes   
{\tt{0}} to {\tt{0}}, then we denote it as {\tt{L}}, while {\tt{1}} to {\tt{0}} is denoted as 
{\tt{R}}.  When we make this notation, the first bit case would 
correspond to {\tt{R}}, since it was from {\tt{1}}, 
rather than {\tt{0}} (at $t=+1$).  The second and third cases would be 
{\tt{L}} and {\tt{R}}, respectively. In Fig. \ref{Irreversible} [i], 
the diagram shows the paths of computation 
for irreversible gates applied to the three bits we just considered.  
If we assume equal probability of initial preparation, 
then the total entropy after $t=+3$ would be $S=k\ln 8$.

Let us now consider a time-reversal process of irreversible gates as a  
nondeterministic computation.  Starting with the process at $t=+3$, 
it is going to send the third bit {\tt{0}} to {\tt{1}} because the 
irreversible gate restored it to {\tt{0}} from {\tt{1}}.  
Thus, nondeterministic computation chooses the acceptable path, 
a reversible process of the irreversible gate, to {\tt{1}}.
Second, in the $t=+2$ case, the reverse process acts 
on the second bit to send {\tt{0}} to {\tt{0}}.  Finally, at $t=+1$, 
it is going to send {\tt{0}} to {\tt{1}} again. 
Similar to the irreversible paths in Fig. \ref{Irreversible} [i], 
the time-reversal nondeterministic computation may be considered (see Fig. \ref{Irreversible} [ii]).

With the recent success in quantum information science, there have been attempts to understand 
or model the universe as a computational process \cite{lloyd1,lloyd2,vedral}.
In particular, in \cite{lloyd2}, the number of computational operations for the universe since 
the big bang has been estimated. 
Similar to the method used in \cite{lloyd2}, 
we make an assumption of the irreversible computations that the 
universe has performed since the big bang. 
Since irreversible computation generates an increase in entropy,
 similar to Fig. \ref{Irreversible} [i], we assume that 
these computations yield the total entropy of the current observable universe; $S=k\ln \Omega$ where 
$\Omega$ is the number of equally accessible microstates of the universe.

With this assumption, 
we now consider nondeterministic computations, i.e.,  the time reversal of the irreversible 
computation with total entropy $S$.   
Similar to Fig. \ref{Irreversible} [ii], we identify $|\theta\rangle_B$ in (\ref{Psi}) 
with the nondeterministic computation that travels backwards in time to the big bang, where 
its time reversal process of irreversible processes generates the entropy $S$. 
Moreover, since $|\theta\rangle_B$ is moving backwards in time, 
it can be considered a Dirac-type negative sea filling 
up the observable universe.  
Therefore, the apparatus state $|\theta\rangle_B$ (or $Q_{\theta}^{B}$) in the two-system in (\ref{Psi}) 
corresponds to the choice of the observable in observing 
the universe, i.e., $\mathcal{O}_{\theta}^{A}$ in (\ref{TheEquation}), which fills up as negative sea.

\section{Remarks}
It is noted that although there are $\Omega$ equally 
probable possible microstates of the observable universe, 
the prepared state corresponds to the observer's 
choice according to (\ref{TheEquation}). 
In \cite{song5,song2}, the concept 
of the subjective universe model has been proposed as a way out of the dilemma of describing consciousness, as 
the observer and the observed cannot be separated. The description provided in this paper suggests that 
the reality should correspond to the observer's choice, therefore, subjective reality.

We now wish to discuss the following four points resulting from the main argument. 
First, in \cite{song5}, after showing the symmetry breakdown 
between the Schr\"odinger and the Heisenberg pictures, a possible 
resolution to the dilemma of describing self-observation of 
the reference frame was studied.
In particular, it was discussed that, based on the postulate that what the observer observes 
is time forwarding, the Heisenberg picture with 
time going backwards should be the appropriate description of nature. 
This conclusion is consistent with our proposed subjective universe model, 
where it is the reference frame going backward 
in time that fills up the observable universe as a negative sea.

Second, in Section 2, we reviewed the observable 
as the exact and full mathematical representation of the observer's conscious state.  
The reasoning behind this identification was as follows:  
Since physics should ultimately provide a description 
between the observing party and the object being observed 
 (Fig. \ref{Subj} [ii]), it was argued that the 
observable should be the final description on the observer's side.
It tried to extend the single-system protocol (Fig. \ref{Dirac} [ii]) 
to the whole universe and argue that just 
as the state vector is as far as one can go in describing the 
object side, the observable is also as far as one can go in providing 
a description of the observing party's side. 
Equating the observable of choice with the apparatus state 
that travels backwards in time, as discussed in this paper, is consistent with  
the observable being the ultimate and exact description of the observer.

Third, the NP computation used in this paper 
was discussed in \cite{song7} in the context of the P versus NP problem. 
This is a well-known problem in computational complexity which asks 
if the class of NP computation, i.e., nondeterministic in polynomial time, 
is equal to the deterministic computation in polynomial time, i.e., P computations. 
Based on the identification of P and NP 
as deterministic and nondeterministic physical processes 
in polynomial time, respectively, it has been shown that the   
observer's choice  between two different paths of computation  
(\ref{nondeterministic}) corresponds to an NP that cannot be contained in P.

Fourth, it is noted that the proposed subjective universe model is different from the 
claim that only consciousness exists, i.e., the whole universe 
exists only within one's consciousness.  
The above model of the subjective universe is not making this claim.

{\it{Acknowledgments}}:  This work was supported by the research grant of the Chungbuk National University in 2015


\end{document}